\documentclass[aps,prl,twocolumn]{revtex4}
\usepackage{graphicx}
\usepackage{amsmath}
\usepackage{amssymb}

\begin{document}
\title{Dynamic wetting with two competing adsorbates}
\author{Christian Gogolin}
\author{Christian Meltzer}
\author{Marvin Willers}
\author{Haye Hinrichsen}
\affiliation{Fakult{\"{a}}t f\"{u}r Physik und Astronomie, Universit\"{a}t W\"{u}rzburg, Am Hubland, 97074 W\"{u}rzburg, Germany}

\begin{abstract}
  We study the dynamic properties of a model for wetting with two competing adsorbates on a planar substrate. The two species of particles have identical properties and repel each other. Starting with a flat interface one observes the formation of homogeneous droplets of the respective type separated by nonwet regions where the interface remains pinned. The wet phase is characterized by slow coarsening of competing droplets. Moreover, in 2+1 dimensions an additional line of continuous phase transition emerges in the bound phase, which separates an unordered phase from an ordered one. The symmetry under interchange of the particle types is spontaneously broken in this region and finite systems exhibit two metastable states, each dominated by one of the species. The critical properties of this transition are analyzed by numeric simulations.
\end{abstract}

\pacs{05.70.Ln, 61.30.Hn, 68.08.Bc}

\maketitle
\parskip 2mm

\section{Introduction}
\label{sec:introduction}

\def\makevector#1{\text{\bf#1}}
\def\xvec{\makevector{x}}

Wetting phenomena are observed in a large variety of situations where an inert surface is exposed to a bulk phase such as a gas or a liquid. Depending on external parameters like chemical potential and temperature the internal forces between the particles and the solid may lead to the formation of a thin layer of a different thermodynamical phase, a so-called wetting layer~\cite{DeGennes}. The morphology and the thickness of the layer depends on how the free-energy contributions at the interfaces between the solid, wetting layer and the bulk phase balance one another. If the bulk phase is thermodynamically favorable the wetting layer remains bound to the surface and is characterized by microscopically finite average width. However, approaching the point where the gas phase and the wetting layer coexist in the bulk, the system undergoes a wetting transition. Beyond this transition the wet phase becomes more favorable in the bulk so that the layer grows, eventually reaching a macroscopic size.

The phase diagram of a system with wetting layers could be rather complex exhibiting a variety of surface phase transitions, prewetting phenomena, and multicritical behavior~\cite{WettingReview,NakanishiFisher}. For example, if the temperature is varied while moving along the coexistence curve of wetting layer and bulk phase, a transition may take place at a temperature $T_W$ beyond which the thickness of the
layer becomes infinite. This transition, known as \textit{continuous wetting}, is usually first order, although in certain models the transition is continuous. On the other hand, varying the chemical potential difference between the two phases and moving towards the coexistence curve at $T > T_W$, a different type of transition takes place, which is referred to as \textit{complete wetting}.

\begin{figure}[t]
\includegraphics[width=240pt]{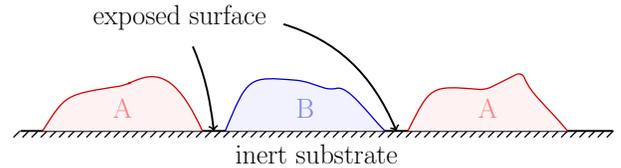}
  \caption{\label{fig:2arsosoverview}(color online) Sketch of a wetting process with competing adsorbates: Two types of particles $A$ and $B$, which strongly repell each other, are adsorbed on top of an inert substrate, forming a wetting layer of mutually avoiding droplets.}
\end{figure}

In many situations it is reasonable to assume that a statistically stationary layer is in thermal equilibrium with its environment. In the 1980's many authors studied such wetting transitions experimentally and theoretically within the framework of equilibrium statistical mechanics  (for a review, see~\cite{WettingReview}). Within this approach, a wetting transition is usually modeled as the unbinding of an interface from a wall. The interface configuration is described by a height function $h(\xvec)$ above the point $\xvec$ on the substrate. The model is then defined in terms of an effective energy functional~\cite{EquilibriumFieldTheory}
\begin{equation}
{\cal E} = \int d^{d}x \biggl[ \frac{\sigma}{2}
(\nabla h)^2 + V\bigl(h(\xvec)\bigr) \biggr],
\end{equation}
where $\sigma$ is the effective surface tension of the interface, $V(h)$ is a potential accounting for the interaction between the wall and the interface, and $d$ is the interface dimension (usually $d=2$). In the nonwet phase the potential $V$ contains an attractive component which binds the interface to the wall. Assuming thermal equilibrium, the probability of finding the interface in a certain configuration is
then given by the canonical distribution
\begin{equation}
\label{EquilibriumEnsemble}
P[h] \sim \exp\bigl(-\beta\,{\cal E}[h]\bigr).
\end{equation}
As the parameters describing the system are varied, the attractive component of the potential may become weaker so that it is no
longer able to bind the interface, leading to a wetting transition.

If one is interested not only in static properties but also in time-dependent features such as dynamical roughening, one usually introduces a stochastic Langevin equation that reproduces the equilibrium distribution~\eqref{EquilibriumEnsemble} in the limit $t \to
\infty$. Assuming short-range interactions and keeping only the most relevant terms in the renormalization group sense, one is led to the Edwards-Wilkinson equation with a potential~\cite{BarabasiStanley}
\begin{equation}
\label{EW}
\frac{\partial h(\xvec,t)}{\partial t} = \sigma\nabla^2 h(\xvec,t) -
\frac{\partial V\bigl(h(\xvec,t)\bigr)}{\partial h(\xvec,t)}
 \noindent + \zeta(\xvec,t)\,,
\end{equation}
where $\zeta(\xvec,t)$ is a zero-average Gaussian noise field with
a variance
\begin{equation}
\label{Noise} \langle\zeta(\xvec,t)\zeta(\xvec',t')\rangle=
2\Gamma\delta^{d-1}(\xvec-\xvec')\delta(t-t')\,,
\end{equation}
and a noise amplitude $\Gamma=k_B\,T$. 

In contrast to the canonical ensemble, the definition of the model as a dynamical process allows one to go beyond the realm of equilibrium thermodynamics and to study the influence of nonequilibrium effects, which manifest themselves as a violation of detailed balance. In fact, during the past decade the study of wetting transitions \emph{far} from equilibrium emerged a s a new sub-topic of the field~\cite{HinrichsenEtAl97,HinrichsenEtAl03,MunozHwa,Wetting2,GiadaMarsili,Candia,MunozWetting,MunozRev,FurtherDevel}. In these theoretical studies it was shown that nonequilibrium wetting transitions may differ significantly from their equilibrium counterparts, exhibiting e.g. different types of critical behavior and new macroscopically observable physical phenomena. 

In the present work we investigate what happens if the surface is exposed to a gas consisting of \textit{two} different types of particles, say $A$ and $B$. Similar situations, but with a focus on catalytic processes, were investigated recently in \cite{Oshanin}. The two species of particles are assumed to repel each other strongly, leading to the formation of competing droplets of either type, as sketched in Fig.~\ref{fig:2arsosoverview}. In our model the repelling force is implemented as a dynamical constraint that deposition is prohibited if it were to cause a direct contact between an $A$ and a $B$ particle. Moreover, the two types of particles are assumed to have identical physical properties, establishing a $A \leftrightarrow B$ symmetry in the model. Certainly it is difficult if not impossible to realize these assumptions experimentally. However, here we are primarily interested in the theoretical question of how such a symmetry influences the universal critical behavior of wetting transitions. For simplicity the present work is restricted to the case of detailed balance, where methods of equilibrium statistical mechanics can be applied. However, the generalization to the nonequilibrium case is straightforward.

The main results of this paper are the following:\\
\indent(i) The competition of $A$ and $B$ islands slows down the dynamics significantly, particularly in the unbound phase, where an extremely slow coarsening process is observed.\\
\indent(ii) In the case of a one-dimensional substrate the phase structure is the same as in the single-species case.\\
\indent(iii) In two (and higher) dimensions an additional transition line divides the bound phase into two parts. \\
\indent(iv) For a growth rate $q=0$ this additional transition is found to belong to the same universality class as the kinetic Ising model. For $q=1$ one observes an unusual type of transition.

\section{Definition of the model}

\begin{figure}[tb]
  \includegraphics[width=80mm]{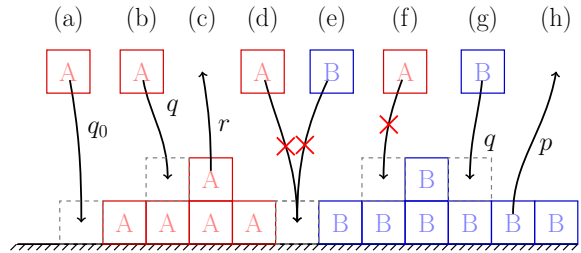}
  \caption{\label{fig:2arsos}(color online) Dynamics of the two-species model. Particles are deposited on the substrate at rate $q_0$ (a) and on islands of the same type at rate $q$ [(b) and (g)]. Moreover, they evaporate from the edges of terraces at rate $r$  (c) and from the middle of plateaus at rate $p$  (h). Particles of different types are not allowed to touch each other [(d) - (f)].}
\end{figure}

\subsection{Single-species model}
Before introducing the two-species model let us briefly recall the single-species model introduced in~\cite{HinrichsenEtAl97}. The single-species model is defined as a solid-on-solid growth process on a flat substrate represented by a $d$-dimensional hypercubic lattice with $N = L^d$ sites. The configuration of the wetting layer is described by an interface without overhangs, meaning that each lattice site $i$ of the substrate is associated with an integer height variable~$h_i$. The interface evolves by deposition and evaporation of particles restricted by two constraints. On the one hand the interface obeys the so-called \textit{restricted solid-on-solid} (RSOS) condition that the heights of neighboring lattice sites $i,j$ may differ by at most one unit:
\begin{equation}
  \label{RSOS}
  |h_i - h_j| \leq 1 .
\end{equation}
This constraint was first introduced in~\cite{kim89} and imposes an effective surface tension. On the other hand, the chemically inert substrate is modeled as a hard-core wall by imposing the dynamical constraint
\begin{equation}
  \label{Wall}
  h_i \geq 0.
\end{equation}
The model evolves by random-sequential dynamics, i.e., a site of the lattice is randomly selected and one of the following processes is carried out, provided that it does not violate the constraints~\eqref{RSOS} and \eqref{Wall}:
(i) deposition on the substrate at rate $q_0$,
(ii) deposition on top of islands at rate $q$,
(iii) evaporation from the middle of plateaus at rate~$p$,
(iv) evaporation at the edges of plateaus at rate~$r$.

If the resulting configuration were to violate the constraints~\eqref{RSOS} or~\eqref{Wall} the attempted move is rejected. As usual, the time scale is fixed by choosing one of the rates, e.g., $r=1$. For $p\neq1$ the process defined above can be shown to violate detailed balance in the stationary state, while for $p=1$ and $q<1$ the stationary state is given by a Boltzmann-Gibbs distribution.

The wetting transition in this model can be understood as follows. Far away from the wall a free interface propagates with a certain average velocity $\frac{d}{dt}\langle h \rangle=v(q,p)$ which depends on the growth rate $q$. Varying the growth rate this velocity changes sign at a well-defined threshold $q_c(p)$. For $v(q,p)<0$ the interface moves backwards until it reaches the substrate  where it continues to fluctuate in a stationary bound state, while for $v(q,p)>0$ the interface detaches from the bottom layer.

\subsection{Two-species model}
In the present work we generalize the model introduced in~\cite{HinrichsenEtAl97} as to describe deposition and evaporation of two different types of particles, labeled by $A$ and $B$. Both species of particles are completely symmetric and obey the same dynamic rules as in the original model. In addition we require that the two species repel each other, implemented by the dynamical constraint that particles of different types are forbidden to be in contact with each other. For example, a particle of type $A$ must not be deposited on top of a $B$ and vice versa. Similarly, particles of different types cannot be deposited at adjacent sites (see Fig.~\ref{fig:2arsos}). Starting with a flat interface at zero height, these dynamical rules ensure that at each lattice site the wetting layer consists of only one type of particles, leading to homogeneous droplets, consisting either of $A$ or $B$ particles. Note that this model is completely symmetric under an interchange $A \leftrightarrow B$, establishing a global $Z_2$ symmetry.

The present study is restricted to the case $p=1$, where stationary bound states obey detailed balance so that methods from equilibrium statistical mechanics can be applied. A full analysis of the genuine nonequilibrium case $p\neq 1$ will be presented elsewhere.

\begin{table}[t]
  \caption{\label{tab:choiceofprocesses} List of the probabilities for deposition and evaporation in a local update, normalized by $n=\max(r + q, 2\,q_0)$.}
  \begin{ruledtabular}
    \begin{tabular}{l@{\extracolsep{\fill}}r|cr|cr|cr}
      Process                           && $h_i=0$  && $h_i>0$   && $h_i<0$    & \\
      \hline
      Deposition of $A$: $h_i\rightarrow h_i+1$ && $q_0/n$ && $q/n$  &&    0       & \\
      Deposition of $B$: $h_i\rightarrow h_i-1$ && $q_0/n$ &&      0     && $q/n$  & \\
      Evaporation: $h_i\rightarrow h_i- \mathrm{sgn}(h_i)$      &&     0     && $r/n$ && $r/n$ & \\
    \end{tabular}
  \end{ruledtabular}
\end{table}

\begin{figure}[t]
  \includegraphics[width=\linewidth]{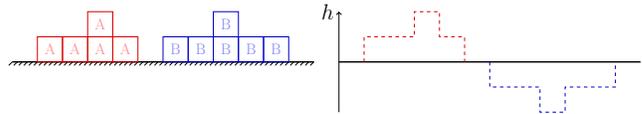} \vspace{-3mm}
  \caption{\label{fig:implementation}(color online) Numerical implementation: Droplets consisting of $A$ and $B$ particles are represented by positive and negative heights, respectively.}
\end{figure}

As sketched in Fig.~\ref{fig:implementation}, the generalized model can be implemented numerically in a very elegant way by representing a wetting layer of type $A$ by positive and a layer of type $B$ by negative heights. As before, $h_i=0$ stands for an unoccupied site where the substrate is exposed to the gas phase. Compliance with the additional rule that particles of different types may not be in contact with each other is then consistently ensured by extending the RSOS constraint~\eqref{RSOS} to negative heights. Using this representation, the generalized model can be simulated by carrying out the following local update rule:\\
\indent 1. A site $i$ is chosen at random.\\
\indent 2. A deposition or evaporation process is selected according to the probabilities listed in Table~\ref{tab:choiceofprocesses}.\\
\indent 3. The attempted update is rejected if the resulting configuration violates the restricted solid--on--solid condition~\eqref{RSOS}, otherwise it is carried out.\\
\indent 4. The time variable is incremented by $\Delta t=\frac{1}{nN}$, where $n=\max(r + q, 2\,q_0)$.

The successive execution of $N$ local updates is referred to as one Monte Carlo step (MCS). As before, the overall time scale can be fixed by choosing one of the rates, e.g. by setting $r=1$. 

The representation of $A$ and $B$ layers by positive and negative heights is not only technically useful. As we will see below, it is also instructive in analytical considerations, as it allows one to relate the two-species model to the original single-species model.

In what follows we are primarily interested in the physically relevant case of a two-dimensional substrate because this is the lowest dimension where the competition of the two particle species leads to an additional phase transition in the pinned region. In all numerical simulations we use periodic boundary conditions in order to minimize finite size effects. 

\section{Phase diagram}
To describe and analyze the model, we mainly consider three different order parameters. The first quantity is the \emph{occupation balance}, defined by
\begin{equation}
  \label{eq:occupationbalance}
  b = \frac{1}{N} \sum_{i=1}^N h_i\,. 
\end{equation}
Since particles of type $A$ are represented by positive and particles of type $B$ by negative occupation numbers $h_i$, the occupation balance is a measure of the surplus of one type of particles and thus indicates a broken $A\leftrightarrow B$ symmetry.

The second quantity is the \emph{density of unoccupied sites}
\begin{equation}
  \label{equ:densofuols}
  \rho^{(0)} = \frac{N^{(0)}}{N}  = \frac{1}{N} \sum_{i=1}^N \delta_{h_i,0} \,.
\end{equation}
This parameter indicates how strongly the interface is pinned to the bottom layer. In particular, for large deposition rates, where the surface is covered by competing islands which are separated by lines of unoccupied sites, this order parameter is a measure of the length of the perimeter of the islands and thus allows us to draw conclusions about their size and the roughness of their borders.

\begin{figure}[t]
  \includegraphics{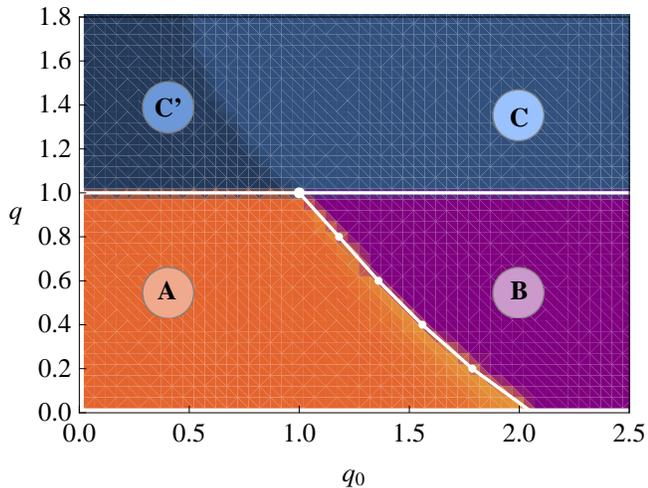}
  \caption{\label{fig:phasediagram}(color online) Phase diagram of the two-species model for $p=1$ in 2+1 dimensions, comprising a symmetric phase (A), a symmetry-broken phase (B), and a rough phase (C/C').}
\end{figure}

\begin{figure*}[t]
  \includegraphics[width=179mm]{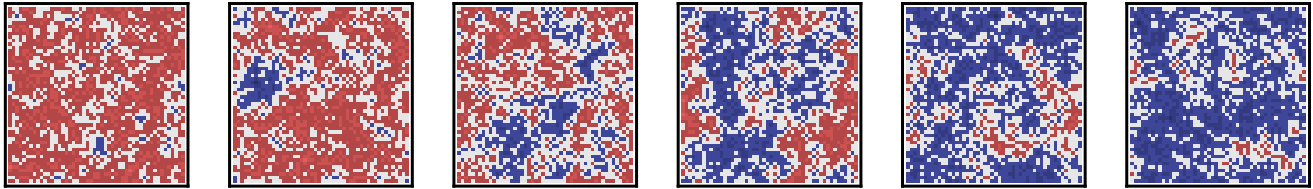}
  \caption{\label{fig:snapshots}(color online) Six snapshots of a $50 \times 50$ Lattice during a flipping process with $q_0=1.17$ and $q = 0.82$ from a state dominated by red (bright) to a state dominated by blue (dark) particles. The brightness indicates the height, white stands for unoccupied sites. }
\end{figure*}

Finally, we are also interested in the \emph{interface width}
\begin{equation}
  w \;=\; \sqrt{\frac{1}{N}\sum_i|h_i|^2-\Bigl(\frac{1}{N}\sum_i|h_i|\Bigr)^2}
\end{equation}
which quantifies the roughness of the interface. These order parameters allow one to identify the following phases (see Fig.~\ref{fig:phasediagram}):\\
\indent(A) \textit{Symmetric phase:} For $q<1$ and small values of~$q_0$ the interface fluctuates close to the wall, forming small short-lived islands of either type. The average occupation balance is zero and the width of the interface saturates at a finite value.\\
\indent(B) \textit{Symmetry-broken phase:} For large values of $q_0$ and $q<1$ one observes coarsening patterns of competing islands. The islands have a pancake like shape and the interface width stays finite. Finite systems have two metastable states, that each feature a surplus of one of the particle types and become stable in the thermodynamic limit. The model has a stationary symmetry-broken state with a nonzero occupation balance. This phase does not exist in 1+1 dimensions.\\
\indent(C,C') \textit{Rough phase:} As free parts of the interface tend to grow for $q>1$, this phase is characterized by the competition of islands of the two particle types. As these islands coarsen, one observes a monotonic increase of the average interface width. This coarsening process turns out to be extremely slow.

  In finite systems a state with only one type of particles is reached after a characteristic time scale. In the limit of large systems this time scale almost certainly grows exponentially or super-exponentially with the system size. After the expulsion of one of the particle types the interface detaches from the bottom layer, entering a moving non-stationary state. Note that this behavior differs significantly from the one of the single-species model, where the interface detaches immediately.

The phases (A) and (B) are separated by a phase transition at a well-defined critical threshold. This raises the question whether this line extends to the upper part of the phase diagram, separating the rough phase into two parts (C) and (C'). As will be discussed below, these regions are quite different in character. However, we were not able to locate a transition line between (C) and (C') in a robust and reproducible way. Therefore we believe that the apparent phase boundary for $q>1$ is a precursor of the (A)-(B) transition rather than a line of genuine phase transition.

\section{Classification of the phases by means of the Binder cumulant}

\begin{figure}[t]
  \includegraphics[width=80mm]{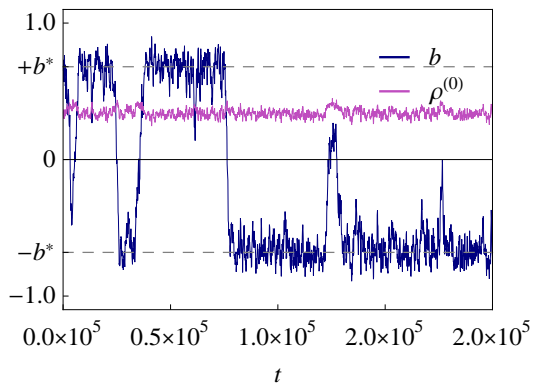}
  \vspace{-3mm}
  \caption{\label{fig:flipdemo}(color online) Occupation balance $b$ and density of unoccupied sites $\rho^{(0)}$ in a system with $50 \times 50$ sites for $q_0=1.17$ and $q=0.82$. Every data point was averaged over 100 MCSs. The system flips between two metastable states. Obviously the flipping is correlated with a small increase of the density of unoccupied sites.}
\end{figure}

\begin{figure}[t]
  \includegraphics[width=85mm]{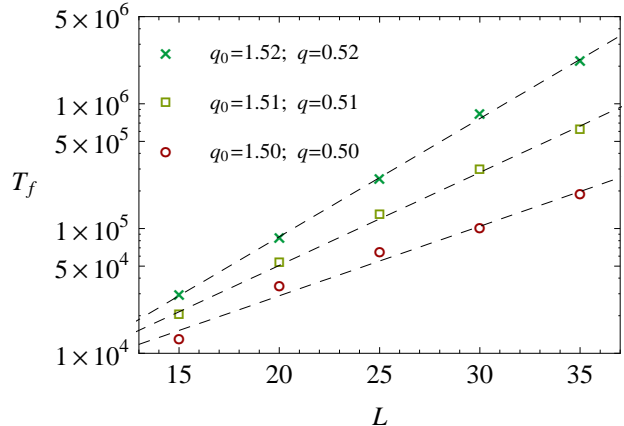}
  \vspace{-2mm}
  \caption{\label{fig:flippingtime}(color online) The characteristic time interval $T_f$ between two flipping processes depends exponentially on the system length $L$. The two metastable states of the symmetry--breaking phase become stable in the thermodynamic limit. In 2+1 dimensions the two-species model exhibits spontaneous symmetry-breaking.}
\end{figure}

In phase (B) one observes coarsening and pancakelike islands which compete one another.
Finite systems have two metastable states, which are each dominated by one of the particle types.
This leads to a spontaneous breaking of the $A\leftrightarrow B$ symmetry, in a similar way as in ferromagnets. 
In \textit{finite} systems the symmetry-broken state is only metastable against fluctuations. 
Therefore one observes occasional flips from an $A$-dominated to a $B$-dominated state and back. Fig.~\ref{fig:snapshots} shows a sequence of six snapshots of such a flip extending over 4000 Monte Carlo updates. As can be seen, a flip occurs whenever minority islands generated by fluctuations reach a critical size that suffices to displace the majority type.

The repeated flipping in finite systems can be observed by monitoring the occupation balance $b$. A typical plot of the behavior in this phase is shown in Fig.~\ref{fig:flipdemo}. As can be seen, the occupation balance $b$ fluctuates around one of two values $\pm b^*$ which characterize the two metastable states. This behavior is intermitted by flipping processes from one of the states to the other. As shown in Fig.~\ref{fig:flippingtime} the average time between two flipping processes $T_f$ grows exponentially with the system size. Thus in the limit $L\to\infty$ the model has two thermodynamically stable ground states that spontaneously break the symmetry under interchange of the two particle types.

Fig.~\ref{fig:flipdemo} reveals another interesting fact: Obviously, the flipping is weakly correlated with a transient increase of the density of unoccupied sites. To understand this increase qualitatively let us again consider the snapshots shown in Fig.~\ref{fig:snapshots}. While the system is in one of the metastable states the number of unoccupied sites is correlated with the number and the size of the minority islands, as most unoccupied sites are arranged along their perimeter. However, during the flipping the system is dominated by two big islands, one of each particle type, that are separated by a fissured, but preferentially straight, domain wall, along which the unoccupied sites are located. For this, the number of exposed sites (gray pixels) is maximal during the flipping process (see Fig.~\ref{fig:snapshots}).

The two phases are associated with different characteristic probability distributions of the occupation balance. In the symmetric phase (A) a finite system exhibits a Gaussian distribution centered around zero, while in the symmetry-broken phase (B) the distribution shows two peaks localized at $\pm b^*$. This qualitative difference can be studied by measuring the Binder cumulant~\cite{binder81}
\begin{equation}
  \label{eq:momentsparameter}
  U = 1 - \frac{1}{3}\,\frac{\langle b^4 \rangle^{\ }}{\langle b^2 \rangle ^2}.
\end{equation}
In the symmetric phase, where $b$ is normally distributed, the cumulant vanishes. In the symmetry-broken phase, where the occupation balance fluctuates only slightly around $\pm b^*$, the Binder cumulant should tend to the value
\begin{equation}
  \label{eq:approxiamtionofcflipping}
  U \approx 1 - \frac{1}{3} \frac{(b^*)^4}{\left[ (b^*)^2 \right]^2} = \frac{2}{3} .
\end{equation}
Interestingly the Binder Cumulant can also be used to identify regions in which the interface is propagating at a constant velocity $v$. Such regions can be expected to exist in the region $q>1$. In this case $b_t \approx \pm v\,t$, such that one finds
\begin{equation}
  \label{eq:approxiamtionofcgrowing}
  U \approx 1 - \frac{1}{3} \frac{1/T\,\int_0^T (v\,t)^4 \, \mathrm{d} t^{\ }}{\left[ 1/T\,\int_0^T (v\,t)^2 \, \mathrm{d} t \right]^2} = \frac{2}{5}\,.
\end{equation}
%

\begin{figure}[t]
  \includegraphics[width=76mm]{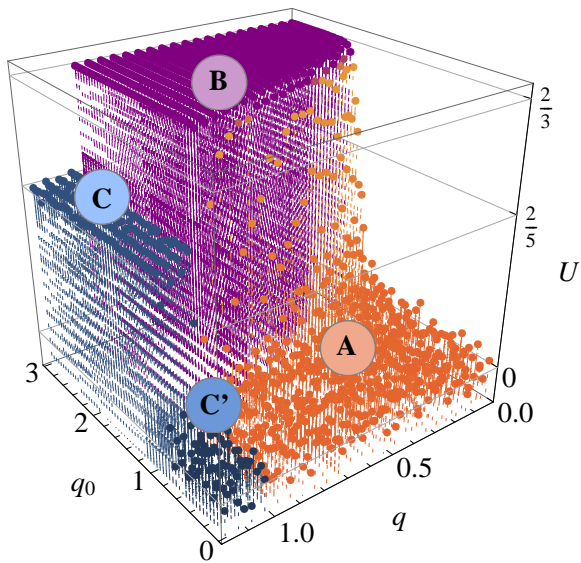}
  \caption{\label{fig:phasediagram2}(color online) Binder cumulant $U$ as a function of the parameters $q$ and $q_0$ in a 2+1-dimensional system with $N=50^2$ sites averaged over $T=2\times10^5$ MCSs after an equally long relaxation time. The same data are shown as a density plot in the background of Fig.~\ref{fig:phasediagram}. Orange (A) corresponds to the symmetric phase, violet (B) to the symmetry-broken phase and blue (C and C') to the growing phase.}
\end{figure}

As demonstrated in Fig.~\ref{fig:phasediagram2} these values are nicely reproduced in numerical simulations and can be used to distinguish between the phases (A) and (B). For $q>1$, the interface of a \textit{finite} system detaches from the bottom layer after a characteristic time $T_d$. In region (C) this time is much smaller than the simulation time such that the Binder cumulant takes the value from Eq.~\eqref{eq:approxiamtionofcgrowing}. In region (C'), however, the cumulant remains close to zero. Note that this sudden change in the Binder cumulant between (C) and (C') does not necessarily mean that a phase transition takes place, rather it simply indicates that the coarsening process in region (C') is so slow that the actual simulation time does not suffice to observe the detachment of the interface.

\section{Stationary equilibrium state}
\subsection{Detailed balance}
For $p=1$ and $q<1$ the model is in the bound phase and evolves towards a fluctuating stationary state. As will be shown below, this stationary state obeys detailed balance so that methods of equilibrium statistical mechanics can be applied. More specifically, it turns out that the stationary state is characterized by a canonical distribution, i.e., the probability of finding an interface configuration $\{h_i\}_{i=1}^N$ which is consistent with the RSOS condition~\eqref{RSOS} is given by
\begin{equation}
  \label{eq:probability}
  P(\{h_i\}_{i=1}^N) = \frac{1}{Z_N} \exp[-\mathcal{H}(\{h_i\}_{i=1}^N)] \,.
\end{equation}
Here $\mathcal{H}(\{h_i\}_{i=1}^N)$ is an energy functional and
\begin{equation}
  Z_N = \sum_{\{h_i\}_{i=1}^N} \exp[-\mathcal{H}(\{h_i\}_{i=1}^N)]
\end{equation}
is the partition sum over all configurations $\{h_i\}_{i=1}^N$ obeying the RSOS constraint. The energy functional is of the form
\begin{equation}
  \label{eq:energyfunctional}
  \mathcal{H}(\{h_i\}_{i=1}^N) = \sum_{i=1}^N V(h_i)
\end{equation}
meaning that it associates with every height value $h_i$ a potential energy $V(h_i)$ that depends on the rates $q_0$ and~$q$. Following~\cite{HinrichsenEtAl03} this potential is given by
\begin{equation}
  \label{eq:potential}
  V(h) =
  \begin{cases}
    - \ln(q/q_0) & h = 0 \\
    - |h| \ln(q) & h \neq 0\\
  \end{cases} .
\end{equation}
Note that this potential is symmetric under the change of sign $h\to-h$, reflecting the $A\leftrightarrow B$ symmetry. Inserting Eqs.~\eqref{eq:energyfunctional} and \eqref{eq:potential} into Eq.~\eqref{eq:probability} yields the distribution
\begin{equation}
  \label{eq:statdist}
  \begin{split}
    P(\{h_i\}_{i=1}^N) = \frac{1}{Z_N} q^{(\sum_{i=1}^N |h_i|)}\ (q/q_0)^{(\sum_{i=1}^N \delta_{h_i,0})} .
  \end{split}
\end{equation}
Following~\cite{HinrichsenEtAl03} one can easily show that the distribution~\eqref{eq:statdist} is in fact stationary and obeys detailed balance. According to Table~\ref{tab:choiceofprocesses} the deposition of a particle on the substrate or on top of an island changes the probability of a configuration by a factor of $q_0$ or $q$, respectively. Likewise evaporation contributes a factor $1/q_0$ or $1/q$. As the rates for deposition are $q_0$ and $q$ and the rate for evaporation is $1$, this state obeys detailed balance and is therefore stationary provided that $q<1$. 

\subsection{Origin of the phase transition}
Regarding Eq.~\eqref{eq:statdist}, the existence of a phase transition seems to be surprising as there is no explicit interaction, not even a short ranged one, between the sites of the lattice. In fact, the only mechanism that leads to correlations between the sites and therewith to a phase transition in this model is the RSOS constraint.

To understand this mechanism, it is instructive to consider the simple case $q=0$, where the dynamics is restricted to a single monolayer, i.e. $h_i \in \{0,\pm1\}$. In this case the potential~\eqref{eq:potential} is given by 
\begin{equation}
  V(h) =
  \begin{cases}
    0 & h = 0\\
    -\ln(q_0) & h \neq 0
  \end{cases}
\end{equation}
and the probability for finding a configuration with $h_i \in \{0,\pm1\}$ that is consistent with the RSOS constraint~\eqref{RSOS} reads
\begin{equation}
  \begin{split}
    P(\{h_i\}_{i=1}^N) &= \frac{1}{Z_N} q_0^{(\sum_{i=1}^N |h_i|)} \\
    &= \frac{1}{Z_N} q_0^{N^{(+)} + N^{(-)}} = \frac{1}{Z_N} q_0^{N-N^{(0)}}.
  \end{split}
\end{equation}
Here $N^{(\pm)}$ denotes the total number of sites with $h_i=\pm1$, so that $N = N^{(+)} + N^{(-)} + N^{(0)}$. The occupation balance is then given by $b = \frac{1}{N}(N^{(+)} - N^{(-)})$. Obviously, for small $q_0$ the system prefers sparsely occupied configurations whereas for large $q_0$ configurations with a high density of islands become more likely. As the probability for a \textit{specific} configuration depends only on the number of occupied states $N^{(+)} + N^{(-)}$ the behavior of the system is determined by the number of possible configurations $m(N^{(+)},N^{(-)})$ for given particle numbers $N^{(+)}$ and $N^{(-)}$. This multiplicity of states plays a key role for the phase transition between phases (A) and (B).

Without the RSOS constraint the multiplicity would, regardless of the dimensionality of the lattice, just be given by
\begin{equation}
  m(N^{(+)},N^{(-)}) = \frac{N!}{N^{(+)}!\,N^{(-)}!\,N^{(0)}!} .
\end{equation}
Therefore, without the RSOS condition, the most probable configurations would be characterized by $N^{(+)} \approx N^{(-)}$, irrespective of the value of $q_0$. 

\begin{figure}[t]
  \includegraphics[width=\linewidth]{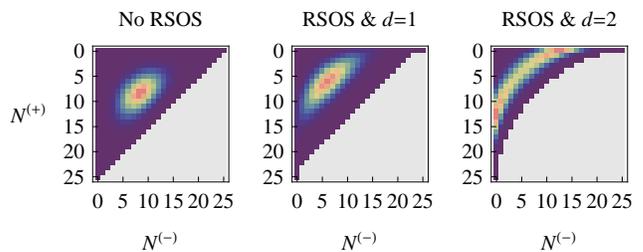}
  \vspace{-3mm}
  \caption{\label{fig:multiplicity}(color online) Multiplicity of states for all possible values of $N^{(+)}$ and $N^{(-)}$ of systems with 25 sites. Light Gray stands for multiplicity zero.}
\end{figure}

The influence of the RSOS constraint on the multiplicities $m(N^{(+)},N^{(-)})$ depends strongly on the dimensionality of the lattice. This is demonstrated in Fig.~\ref{fig:multiplicity}, where the multiplicities for systems with $N=25$ sites are compared for the unrestricted and the restricted case in 1+1 and 2+1 dimensions. As can be seen, the RSOS condition reduces the area where $m$ is nonzero since islands of different types have to be separated by unoccupied sites at zero height. On the other hand, in 2+1 dimensions the influence of the RSOS constraint is so strong that the maxima are shifted away from the diagonal to the edges of the diagram. Therefore, if $q_0$ is larger than some critical value $q_0^C$, the system is preferentially driven into one of these configurations at the edges, explaining why the system prefers a symmetry-broken state. Fig.~\ref{fig:multiplicity} also demonstrates why the symmetry breaking is not observed in 1+1 dimensions, where the influence of the RSOS constraint is apparently not strong enough to shift the maxima of the multiplicities to the edges.

For $q<1$ higher layers become accessible but they are occupied only rarely so that the islands retain their pancakelike shape. Therefore the situation can be expected to be qualitatively similar. Nevertheless, particles in higher layers tend to stabilize the particles in the layers below, increasing the effective growth rate. Therefore the critical value $q_0^C$ should decreases with increasing $q$. As can be seen in the phase diagram (Fig.~\ref{fig:phasediagram}), this is indeed what the numerical simulations show.

\section{Critical properties}
In this section we study the dynamical critical behavior along the transition line between the phases (A) and (B) and at its upper terminal point. Here the two order parameters defined above require a separate treatment with different initial conditions. For the occupation balance $b_t$ one has to start with a monolayer of only one species of particles, say $A$-particles so that $b_0=1$. With this initial condition the occupation balance $b_t$ is expected to decay at criticality as
\begin{equation}
  \label{eq:expdelta}
  b_t \sim t^{-\delta}.
\end{equation}
On the other hand, the density of exposed sites $\rho^{(0)}_t$ requires to start with a flat interface so that $\rho^{(0)}_0 = 1$. At criticality this quantity is expected to decay as
\begin{equation}
  \rho^{(0)}_t - \rho^{(0)}_\infty \sim t^{-\alpha}
\end{equation}
while the occupation balance vanishes for all $t$. 

Moving away from the critical line into the symmetry-broken phase (B) the occupation balance becomes nonzero and flips between the values $\pm b^*$.
One can therefore use 
\begin{equation}
  b^* = b^*(q_0,q) = \langle | b | \rangle ,
\end{equation}
as a magnetizationlike order parameter to characterize the transition between the two phases. Suppressing flips by choosing sufficiently large system sizes and varying $q_0$ while keeping $q<1$ fixed $b^*$ is found to increase as
\begin{equation}
  \label{eq:magnetisationpowerlaw}
  b^*(q_0,q) \sim [q_0-q_0^c(q)]^\beta .
\end{equation}

By seeking for power laws and by measuring the order parameter $b^*$ we first determined various critical points $q_0^c$ which are listed in Table~\ref{tab:numericvaluesofthephaseboundary}.

\subsection{Transition for $q=0$}
Measuring the dynamic critical exponents $\alpha$ and $\delta$ at the lower end of the transition line $q_0=q_0^c(0),\,q=0$ we find  
\begin{equation}
  \delta = 0.058(4) \quad \text{and} \quad \alpha = 0.5(1) .
\end{equation}
The density of exposed sites relaxes vs the value $\rho^{(0)}_\infty = 0.384(5)$. Moving into the ordered phase (B) with $q=0$ fixed the magnetizationlike order parameter $b^*$ obeys Eq.~\eqref{eq:magnetisationpowerlaw} with a critical exponent of
\begin{equation}
  \beta = 0.125(5) .
\end{equation}

This suggests that at this point the transitions belongs to the universality class of the two-dimensional kinetic Ising model with heat bath dynamics, which is characterized by the exponents $\beta=1/8$, $z\approx 2.125$, and $\alpha=\beta/z \approx 0.0588$~\cite{Ising}. The ordered phase of the kinetic Ising model is known to be characterized by coarsening domains with an average size growing as $t^{1/2}$~\cite{bray94}. One would therefore expect that the density of exposed sites $\rho^{(0)}_t$, which in the present case is a measure of the density of domain walls, decays as $t^{-1/2}$, which is consistent with our results.

This result seems to be reasonable, as at $q=0$, where the height values are restricted to $h_i \in \{0,\pm1\}$, our model is very similar to a kinetic Ising model. The behavior is governed by the competition of bulk noise, caused by evaporation and deposition, and the ordering influence of the RSOS constraint. The rate $q_0$ acts as an effective inverse temperature. Similar behavior was observed in other binary mixture models on two-dimensional lattices \cite{Oshanin}.

\subsection{Transition line for $0<q<1$}
For $0<q<1$ the coarsening islands retain a pancakelike shape since higher layers are exponentially suppressed, as can be seen from the stationary equilibrium distribution~\eqref{eq:statdist}. This means, that the third dimension (height) is switched off so that the resulting process is still effectively two-dimensional. One might therefore expect that the transitions along the entire line between the phases (A) and (B), except for the upper terminal point, belong to the universality class of the two-dimensional kinetic Ising model.

Our simulations suggest that this is not the case, but rather that all three exponents vary continuously when $q$ is increased. The determination of critical exponents is always a delicate issue, especially when, as in the case of $\alpha$ and $\beta$, the power laws contain additional unknown quantities. Being skeptical of these results we have carefully estimated the error bars. The results of our analysis are shown in Table~\ref{tab:numericvaluesofthephaseboundary}.

\begin{figure}[t]
\includegraphics{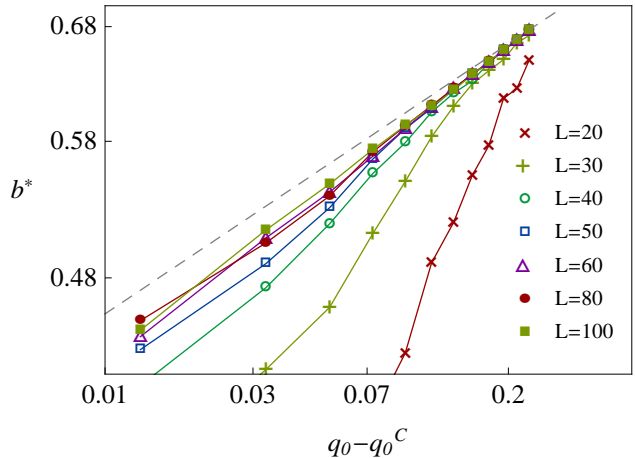}
  \vspace{-2mm}
  \caption{\label{fig:finitesize2}(color online) Finite size scaling at $q=0.2$ to support the hypothesis that the transition is nonuniversal. The dashed gray line corresponds to a slope of $\beta=1/8$.}
\end{figure}

\begin{table}
  \caption{\label{tab:numericvaluesofthephaseboundary}Numerical values for the critical points along the phase boundary $q_0^c(q)$ between the unordered phase (A) and the symmetry-broken phase (B) and various critical exponents $\alpha$, $\delta$, and $\beta$ along that line obtained from simulations of systems with $100\times100$ sites.}
  \begin{ruledtabular}
    \begin{tabular}{l|llllll}
      $q$         & 0       & 0.2     & 0.4     & 0.6     & 0.8     & 1 \\
      \hline
      $q_0^c(q)$ & 2.062(5) & 1.787(5) & 1.560(5) & 1.360(5) & 1.181(5) & 1 \\ 
      $\alpha$  & 0.5(1)   & 0.5(2)   &  0.5(2)  & 0.6(2)   & 0.6(2)   & - \\ 
      $\delta$  & 0.058(5) & 0.063(5) & 0.068(5) & 0.074(5) & 0.098(5) & - \\ 
      $\beta$   & 0.125(5) & 0.14(1)  & 0.16(2)  & 0.19(3)  & 0.21(3)  & - \\ 
    \end{tabular}
  \end{ruledtabular}
\end{table}

\begin{figure}[t]
\includegraphics{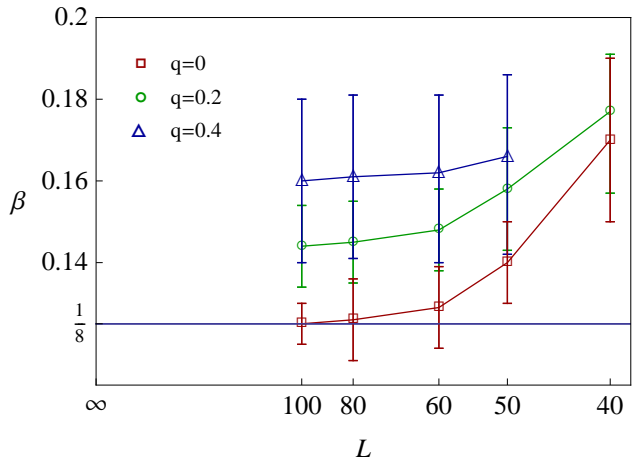}
  \vspace{-2mm}
  \caption{\label{fig:finitesize3}(color online) Magnetization exponent $\beta$ at $q=0$, $q=0.2$, and $q=0.4$ for various system sizes obtained from finite size simulations. For system sizes $L\geq60$ the main error source is the uncertainty in the position of the phase boundary $q_0^c(q)$.}
\end{figure}

The critical points listed in Table~\ref{tab:numericvaluesofthephaseboundary} were estimated in two ways: first by searching for a power law decay of the form Eq.~\eqref{eq:expdelta} while varying $q_0$ for fixed values of $q$. Second by fitting a power law of the form \eqref{eq:magnetisationpowerlaw} to data obtained by stepwise increasing $q_0$ and measuring $b^*$ after sufficiently long relaxation times. Both methods yield consistent results near the lower and upper end of the transition line. In the middle, i.e., around $q \approx 0.5$, the $q_0^c(q)$ values obtained by the second method are slightly smaller. We consider the first method to be more reliable as it is less affected by finite size effects.

The exponent $\delta$ can be determined with high accuracy although its value is fairly small. It is the most reliable indicator for a continuous variation of the exponents along the transition line.

Concerning the exponent $\alpha$ the by far dominating source of errors is the determination of the asymptotic value of the density of exposed sites, which seems to take the value $\rho^{(0)}_\infty = 0.38(1)$ along the entire transition line. Although we see a systematic dependency of the exponent $\alpha$ on the rate $q$ this dependency is not significant.

The most problematic exponent is $\beta$. First, the uncertainty in the position of the phase boundary $q_0^C(q)$ limits the reachable accuracy. In addition, when $q$ is increased the correlation length in phase (B) grows, increasing the influence of finite size effects. The feasible system sizes are in turn limited by the necessity for very long relaxation times. In order to handle both types of errors we performed additional control simulations with stepwise decreased deposition rate $q_0$ and simulations with different system sizes $L \leq 100$.
The results of our finite size simulations are shown in Fig.~\ref{fig:finitesize2} and Fig.~\ref{fig:finitesize3}. The estimated exponent decreases with increasing system sizes, but seem to converge vs a value that is noticeably higher that $1/8$ for $q>0$ in the limit $L \to \infty$.

\subsection{The non-equilibrium phase $q>1$}
For $q>1$ the model has no stationary state and therefore becomes sensitive to the initial conditions. For example, if the substrate is initially covered with several layers of particles of one type, particles of the other type are unable to attach so that the interface grows at constant velocity, just as in the single-species case. However, starting with a flat interface at zero height, the situation is totally different. Here one observes a slow coarsening of competing three-dimensional islands of particles of both types.

As for $q>1$ the deposition rate is higher than the evaporation rate islands of both types tend to grow. The RSOS constraint, however, does not allow slopes greater than one and requires a line of unoccupied sites in between islands of different types. The islands therefore have a pyramidal shape. There is a surface-tension-like effect, that makes large islands grow at the expense of small ones, but the coarsening in this phase is extremely slow. This is due to the fact that the borders of the islands are almost completely immobilized, as the particles in the bottom layer can evaporate only, if beforehand all the particles along the slope of the pyramidal island have evaporated.

\begin{figure}[tb]
  \includegraphics{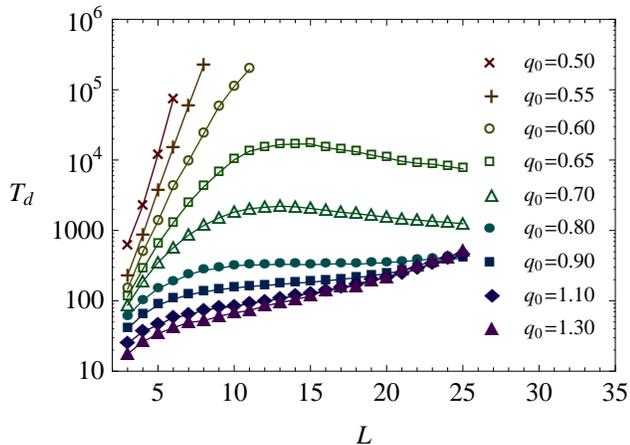}
  \caption{\label{fig:detachtime}(color online) Dependency of the displacement time $T_d$ on the rate $q_0$ and the system size $L$ for $q=1.5$.}
\end{figure}

To quantify the slowdown of the dynamics we performed finite-size simulations with periodic boundary conditions. In these simulations, because of the finite system size, the coarsening process eventually displaces one of the particle types, allowing the interface to detach from the bottom layer and to propagate at constant velocity. As shown in Fig.~\ref{fig:detachtime}, the mean detachment time $T_d$ depends on the rates $q_0$ and $q$ and the system size $L$. The results suggest that the detachment time $T_d$ grows exponentially with the system size $L$, or even faster.

Moreover, the simulations confirm the existence of two different regions (C) and (C') in Fig.~\ref{fig:phasediagram2} for $q>1$, as already observed when measuring the Binder cumulant. In region (C) the formation of islands and the coarsening process start immediately, while in region (C') one observes a low density of small short-lived islands for a long time until one or several large islands, that were generated by fluctuations, begin to grow. 

The different phenomenological properties in these regions can be explained as follows. For $q>1$ the clusters grow by an influx of particles proportional to their area. If $q_0<1$, however, there is also a competing process: the tendency of particles to evaporate from the bottom layer. As such particles are preferentially located at the border of island the strength of this effect is proportional to the perimeter of an island. If $q_0$ is small enough the balance of these two processes defines a critical droplet size from where on islands start to grow. Although the regions (C) and (C') seem to be well separated, we do not see evidence for a phase transition.

For small system sizes $L$ the qualitative different behavior in these two regions even leads to quantitative influence on the detach time (see Fig.~\ref{fig:detachtime}). For high values of $q_0$, i.e. in region (C), the asymptotically exponential increase of the detachment time sets in already for small $L$. For small $q_0$ in region (C') the curves increase much faster for small $L$, leading to detachment times that are several orders of magnitude larger. Surprisingly the curves reach a local maximum followed by a decline and an eventual crossover to the asymptotic exponential increase.

This behavior may be explained as follows: If the system size is smaller than the critical droplet size even subcritical droplets can make the interface detach due to the periodic boundary conditions. The local maximum is reached when the critical droplet size  approximately equals the system size. Increasing the system size further, the probability to generate a critical droplet grows, thereby reducing the detachment time. Here the dynamics is still governed by a single droplet which exceeds the critical size and then grows rapidly until the interface detaches. In much larger systems, however, it is likely that two or more supercritical droplets of different types are created before the first can spread out over the whole substrate. Then the limiting effect is the competition between the pyramidal islands which explains the asymptotic exponential increase.

\subsection{The line $q=1$}
Let us first consider the upper terminal point of the phase transition line located at $q_0=q=1$. Using again the mapping of the model to a growth process with positive and negative heights (see Fig.~\ref{fig:implementation}) it is easy to see that in this case the potential (\ref{eq:potential}) is translationally invariant in the heights so that the dynamics of the model is equivalent to a freely evolving RSOS interface with average velocity zero without the hardcore wall at $h=0$. At this point the surface does no more approach an equilibrium configuration, but rather it continues to roughen with a width increasing as $\sqrt{t}$. In fact, the density of unoccupied sites, that in this case is equivalent to the density of the roughening RSOS interface cut horizontally in the middle, decays like
\begin{equation}
  \langle \rho^{(0)} \rangle \sim t^{-0.50(1)} .
\end{equation}
Moreover, in a finite system starting at zero height the occupation balance is found to perform a random walk, hence
\begin{equation}
  \langle |b| \rangle = b^* \sim t^{0.500(5)} \qquad \text{for}\ (d=2, q=q_0=1).
\end{equation}
For $q_0 > 1$ translational invariance in the heights is broken by a repulsive potential well at zero height. In the special case $q=1$ and $q_0\to\infty$ starting with a monolayer of one type of particles, particles of the other type are unable to attach. The bottom layer acts as an effective inert substrate on which particles of one type form clusters in the same way as they do in the RSOS model. As time proceeds, the interface roughens according to the Edwards-Wilkinson universality class. Therefore the width is expected to increase as
\begin{equation}
  w \sim t^{1/4} .
\end{equation}
This is consistent with our numerical observation that
\begin{equation}
  \langle |b|-\mathrm{sgn}(b) \rangle \sim t^{0.25(3)}  \qquad \text{for}\ (d=2, q=1, q_0\to\infty) .
\end{equation}

For $q_0 < 1$ on the other hand one has an attractive potential barrier at zero height. When the line $q=1$ is approached from above the tendency of islands to grow diminishes as $q$ decreases, and therefore the critical droplet size gets larger and finally diverges as $q=1$ is approached. Starting with an empty lattice the model behaves similar as in the unordered phase (A). Even if the system is started with a few monolayers of particles fluctuations soon create holes in the deposit that grow until the substrate is almost completely exposed and the system again ends up in a situation that is similar to the unordered phase.

\section{Conclusions}
In the present work we have investigated how the phase transition in a model for nonequilibrium wetting is affected by introducing two competing adsorbates with identical properties. In this generalized model we have assumed that the two species of particles repel each other strongly. This leads to the formation of mutually repelling droplets, each consisting of only one type of particles. 

In two dimensions we have identified an additional line of second-order phase transitions in the bound phase (see Fig. 4). It separates an ordered phase, where one of the two particle species takes over, from a disordered phase characterized by many small islands of different type. Defining appropriate order parameters and cumulants we have studied this transition in detail by numerical simulations. The origin of the phase transition can be explained in part by analyzing the partition sum in those parts of the phase diagram where detailed balance is valid.

In the limit $q \to 0$, where the wetting process is governed by a monolayer, this transition exhibits the same critical properties as the kinetic Ising model with heat bath dynamics. This is reasonable since in this case this islands are flat and their interior noise caused by evaporation and deposition competes with ordering influence of surface tension at their boundaries. 

For $0<q<1$ higher layers are involved as well, giving the islands a three-dimensional shape. Nevertheless, their thickness should remain finite and of the order $-1/\ln(q)$ so that one would expect them to behave asymptotically in the same way as in the monolayer case, suggesting that the entire curved line of transitions should belong to the Ising universality class. Contrarily, our numerical simulations seem to give evidence for continuously varying exponents. The question whether these varying exponents are genuine or caused by crossover effects is still open. A summary of the critical exponents can be found in Table~\ref{tab:numericvaluesofthephaseboundary}.

As in the single-species case, the model undergoes a transition from a bound to an unbound phase at the threshold $q=1$ (the horizontal line in Fig.~\ref{fig:phasediagram}). However, the critical properties for a system with an initially flat interface along this line are found to be different. In particular, there is the special point $q = q_0 = 1$, at which the two transition lines meet. It separates the line $q=1$ into two segments. For $q_0>1$ the wetting transition is continuous, while for $q_0<1$ it is discontinuous.

The unbound phase can be divided into two different regimes of different coarsening with an approximate borderline which seems to prolongate the curved transition line from the bound phase. Nevertheless, we do not find evidence for a sharp phase transition between these regimes. In both regions the competition of two species slows down the dynamics. 

The study presented here has been primarily of theoretical interest and was motivated by the question how an additional symmetry between competing adsorbates affects the critical properties of a wetting transition. As soon a this symmetry is broken one expects the process to cross over to an effective single-species behavior after sufficiently long time. It would be interesting to study the case $p\neq 1$, where the process is out of equilibrium even in the bound phase. In this case the additional transition would still exist but probably it will no longer belong to the Ising universality class in the limit of small~$q$.

\end{document}